\documentclass[preprint,times]{elsarticle}

\usepackage{amssymb}
\usepackage{amsmath}

\usepackage{physics}
\usepackage{hyperref}
\usepackage{cleveref} 

\usepackage{booktabs}
\usepackage{multirow}

\usepackage{bm} 

\newcommand{\CC}{C\nolinebreak\hspace{-.05em}\raisebox{.4ex}{\tiny\bf +}\nolinebreak\hspace{-.10em}\raisebox{.4ex}{\tiny\bf +}}

\usepackage{orcidlink}

\journal{Computer Physics Communications}

\begin{document}

\begin{frontmatter}

\title{The Software Landscape for the Density Matrix Renormalization Group}

\author[1]{Per Sehlstedt\orcidlink{0009-0008-9361-9377}}
\ead{pers@cs.umu.se}
\corref{cor1}
\cortext[cor1]{Corresponding author.}

\author[2]{Jan Brandejs\orcidlink{0000-0002-2107-3095}}
\author[1]{Paolo Bientinesi\orcidlink{0000-0002-4972-7097}}
\author[1]{Lars Karlsson\orcidlink{0000-0002-4675-7434}}

\affiliation[1]{organization={Department of Computing Science, Umeå University},
            city={Umeå},
            postcode={901 87},
            country={Sweden}}

\affiliation[2]{organization={Laboratoire de Chimie et Physique Quantiques, CNRS},
            city={Toulouse},
            postcode={310 62},
            country={France}}

\begin{abstract}
The density matrix renormalization group (DMRG) algorithm is a cornerstone computational method for studying quantum many-body systems, renowned for its accuracy and adaptability. 
Despite DMRG’s broad applicability across fields such as materials science, quantum chemistry, and quantum computing, numerous independent implementations have been developed. 
This survey maps the rapidly expanding DMRG software landscape, providing a comprehensive comparison of features among 37 existing packages.
We found significant overlap in features among the packages when comparing key aspects, such as parallelism strategies for high-performance computing and symmetry-adapted formulations that enhance efficiency. 
This overlap suggests opportunities for modularization of common operations, including tensor operations, symmetry representations, and eigensolvers, as the packages are mostly independent and share few third-party library dependencies where functionality is factored out.
More widespread modularization and standardization would result in reduced duplication of efforts and improved interoperability.
We believe that the proliferation of packages and the current lack of standard interfaces and modularity are more social than technical.
We aim to raise awareness of existing packages, guide researchers in finding a suitable package for their needs, and help developers identify opportunities for collaboration, modularity standardization, and optimization.
Ultimately, this work emphasizes the value of greater cohesion and modularity, which would benefit DMRG software, allowing these powerful algorithms to tackle more complex and ambitious problems.
\end{abstract}



\end{frontmatter}

\section{Introduction}
\label{sec: Introduction}

Numerical software is central to modern scientific discovery, enabling breakthroughs across disciplines. This survey examines the diverse and rapidly expanding software landscape for the density matrix renormalization group (DMRG). We provide a map of existing packages and their functionalities that can serve as a valuable resource for researchers and developers interested in the topic. We encourage greater community collaboration to build a more unified foundation based on a well-structured modular hierarchy, allowing everyone to benefit from the full potential of the algorithm and its various extensions.

Tensor network algorithms have become indispensable for modern computational science and engineering, providing powerful frameworks for efficiently representing and simulating quantum many-body systems~\cite{Ors2014oct, Ors2014nov, Ors2019, Okunishi2022}. The modern tensor-based DMRG formulation~\cite{Schollwck2011, stlund1995, Rommer1997, Dukelsky1998} is one of the most successful examples. Its accuracy and versatility have established DMRG as a cornerstone method, inspiring numerous adaptations and extensions far beyond its original scope~\cite{Schollwck2005, Hallberg2006, Schollwck2011}. These advancements have been applied to address problems across various research fields, including quantum chemistry~\cite{White1999}, materials science~\cite{Bauer2020}, quantum computing~\cite{Ayral2023}, and machine learning~\cite{Stoudenmire2016NIPS, Novikov2016Machine, Huggins2019}, among others, covering both one and higher dimensions~\cite{Stoudenmire2012}.

The sweeping variational optimization framework encompassing the various DMRG flavors remains relevant and widely used today. Furthermore, the growing interest in simulating larger and more complex systems, coupled with the algorithm's significant time and space complexity, induces a need to embrace parallelism and high-performance computing (HPC) platforms~\cite{Tian2023, Ganahl2023}. Additionally, exploiting inherent symmetries in the systems is critical for improving DMRG's performance, as it introduces a block-sparsity structure that eliminates redundant computations, accelerates convergence, and enhances numerical stability and accuracy. Symmetry-adapted DMRG versions can yield significant efficiency gains, enabling higher precision and simulations of previously intractable systems~\cite{McCulloch2002, Schollwck2005}.

DMRG's broad applicability has driven the development of numerous software packages. However, researchers often develop these packages independently, motivated by their needs and tailoring them to specific applications. This application-focused and uncoordinated approach has resulted in a lack of standardized modularity. Additionally, it has led to significant duplication of efforts when multiple groups implement similar functionalities. The limited modularity, in particular, poses significant challenges when adapting these codes to HPC platforms. 

This survey aims to map the current software landscape for DMRG, highlighting similarities and differences in features and objectives. Rather than qualitatively grading the packages, we categorize multiple aspects of their functionality. Each package offers unique features and adaptations suited to specific research needs, sometimes extending well beyond DMRG, which we do not capture in the present high-level survey. We primarily strive to inform and assist in navigating the available options without making value judgments. Ultimately, we seek to promote a more unified and structured approach that reduces duplication, enhances modularity, and facilitates advancements that leverage HPC platforms for increasingly ambitious challenges in quantum many-body physics and beyond.

The remainder of this paper is structured as follows: We briefly introduce the DMRG algorithm in \Cref{sec: Foundations and Advancements}. Since we focus on software, we assume the reader is reasonably familiar with the DMRG method. Many excellent reviews are available for those seeking a thorough introduction~\cite{Schollwck2005, Hallberg2006, Schollwck2011, Catarina2023}. We also highlight some developments since its inception, as these have significantly contributed to the proliferation of software packages. 
We describe the methodology of our survey in Section~\ref{sec: Methodology}.
We present all the packages we identified in \Cref{sec: DMRG Software Packages} and compare their features in \Cref{sec: Comparisons}. Finally, we conclude in \Cref{sec: Conclusion}.

\section{Foundations and Advancements}
\label{sec: Foundations and Advancements}

\subsection{The DMRG Algorithm}
\label{ssec: The DMRG Algorithm}

The DMRG algorithm~\cite{White1992, White1993} is a numerical method for approximating dominant eigenvectors of a matrix $H$, particularly the ground state of a quantum many-body system with Hamiltonian $H$. The matrix is often large enough to present significant computational challenges, as the Hamiltonian is defined on the Hilbert space, which grows exponentially with the size of the system. However, the ground state of most quantum many-body systems obeys a boundary law (also known as area law), possibly with a multiplicative logarithmic correction~\cite{Evenbly2014, Srednicki1993, Eisert2006, Hastings2007, Masanes2009, Eisert2010}. Roughly speaking, the boundary law states that a subsystem's entanglement entropy scales with the size of its boundary rather than its volume. This implies that valid ground state candidates are confined to a relatively tiny subspace of the Hilbert space. Ultimately, this allows for data-sparse representations that make computations tractable despite the exponential size of the Hilbert space. 

In the variational principle \cite{Rosenbrock1985}, ground state of a Hamiltonian minimizes the energy expectation value $E$: 
\begin{equation} \label{eq: variational}
    \arg\min_{\ket{\psi}} E(\ket{\psi}) = \arg\min_{\ket{\psi}} \frac{\ev{H}{\psi}}{\braket{\psi}}.
\end{equation}

Originally, DMRG relied on reduced density matrices to identify and track the crucial degrees of freedom defining the system, hence the algorithm's name. However, it was later acknowledged that the singular value decomposition could achieve a similar goal~\cite{Schollwck2011} and that the wave function produced by DMRG naturally takes the form of matrix product states (MPS)
\begin{equation}\label{eq:mps}
|\psi_\mathrm{MPS}\rangle = \sum_{\sigma_1, \ldots, \sigma_L} A^{\sigma_1}_1 A^{\sigma_2}_2 \ldots A^{\sigma_L}_L |\sigma_1 \ldots \sigma_L\rangle,
\end{equation}
where $L$ is the number of sites and each $A^{\sigma_i}_i$ is a matrix of parameters. Single-site space $\sigma_i$ is typically of small dimension, e.g., a chemical orbital of dimension 4.
Since the MPS ansatz is a valid trial function for~\cref{eq: variational}, DMRG is a variational method~\cite{Verstraete2004Density} minimizing $E$ within the given MPS manifold. Minimization is achieved through an iterative sweeping procedure that updates some elements while keeping all others fixed. The update at each sweeping step refines the approximation using an iterative eigensolver such as Lanczos~\cite{Lanczos1950} or Davidson~\cite{Davidson1975}. These projection-based solvers are preferred as they allow for matrix-free implementations that avoid the simultaneous storage of Hamiltonian components in memory. Note that forms of the wavefunction ansatz beyond MPS have been proposed, e.g., T3NS \cite{Gunst2018}.

Typical options for DMRG calculations include distinctions based on system type (infinite versus finite) and boundary conditions (open, periodic~\cite{Pippan2010, Verstraete2004Density}, smooth~\cite{Veki1993, Veki1996}, or infinite~\cite{Phien2012}) tailored to a range of lattice geometries from one-dimensional chains to two-dimensional square \cite{Jeckelmann1998, White2007}, triangular~\cite{Yoshikawa2004, Weng2006}, CAVO~\cite{White1996}, honeycomb~\cite{Jiang2011}, or Kagome~\cite{Sachdev1992, Jiang2008, Yan2011} lattices~\cite{Stoudenmire2012}.

Enhancements such as density matrix perturbation~\cite{White2005}, subspace expansion~\cite{Hubig2015, Dolgov2014jan, Dolgov2014oct}, and the recent controlled bond expansion~\cite{Gleis2023} have notably enabled reliable single-site formulations of DMRG, sometimes delivering significant speed-ups and improved convergence, especially for systems with long-range interactions. DMRG is typically implemented using a two-site formulation, which naturally allows for adaptive precision refinements of $\ket{\psi}$, often helping to avoid convergence issues.
Traditional single-site formulations significantly reduce computation and memory costs but do not allow for such adaptivity, putting the simulation at risk of getting stuck in metastable configurations, i.e., local minima that differ from the ground state~\cite{White2005}. 
These enhancements aim to balance the computational advantages of the single-site approach while reintroducing a degree of adaptivity and mitigating its risks.

Implementations commonly have the capability to exploit symmetries inherent to many quantum models. Symmetry-adapted DMRG can yield significant performance improvements~\cite{McCulloch2002, Schollwck2005}. Incorporating abelian symmetries enables block-sparse representations that reduce computational overhead by eliminating redundant calculations. Furthermore, leveraging non-abelian symmetries can result in even more significant efficiency gains by capturing a richer system structure. However, this entails a significantly increased implementation effort due to the complexity of managing non-commutative group representations, often relying on Clebsch-Gordan coefficients and the Wigner-Eckart theorem to construct and manipulate symmetry-adapted tensor structures properly and efficiently. Despite these challenges, the potential performance improvements offered by non-abelian symmetry-adapted methods make them a powerful tool for tackling complex, highly symmetric systems.

\subsection{Development and Evolution}
\label{ssec: Development and Evolution}

Since its inception in 1992 by Steven R. White~\cite{White1992, White1993}, the DMRG algorithm has continuously developed and evolved to accommodate new applications and system types. It was initially conceived within the framework of renormalization group (RG) methods, building upon earlier ideas in condensed matter physics. However, over time, the language describing DMRG shifted from the RG perspective to tensor networks, particularly matrix product states (MPS)~\cite{Schollwck2011, stlund1995, Rommer1997, Dukelsky1998} and matrix product operators (MPO)~\cite{Verstraete2004, Zwolak2004, McCulloch2007, Pirvu2010, Hubig2017}. The MPS/MPO framework has since become the foundation for understanding and generalizing DMRG.

The applicability of DMRG has spread across several scientific domains. For example, in ab initio quantum chemistry, it has been applied to solve the electronic structure problem in molecules~\cite{Chan2011, Wouters2014Density, Yanai2014, Chan2016, Baiardi2020}. The algorithm's ability to capture static correlations makes it highly effective, particularly in cases where traditional methods fall short. However, unlike traditional methods, DMRG often struggles to capture dynamic correlations. These trade-offs have resulted in hybrid quantum chemistry methods integrating DMRG with traditional multi-reference techniques. For instance, DMRG has been integrated with CASSCF, CASPT2, MRCI, CT, and MRDSRG to combine the methods' strengths and effectively address static and dynamic electron correlations.

DMRG has also been extended to tackle dynamic behavior and properties~\cite{Paeckel2019, Li2020}. Time-dependent DMRG methods, including time-evolving block-decimation (TEBD) \cite{AJDaley2004, Vidal2003, Vidal2004, White2004}, the time-dependent variational principle (TDVP)~\cite{Haegeman2011, Milsted2013, Haegeman2016, Kloss2018}, local Krylov~\cite{Schmitteckert2004, Feiguin2005} and global Krylov~\cite{GarcaRipoll2006, Dargel2012} techniques, and MPO $W^{\mathrm{I, II}}$ methods~\cite{Zaletel2015}, have opened pathways for real- and imaginary-time evolution studies. 
Additionally, DMRG has been extended for excited-state calculations~\cite{Chandross1999, Khemani2016}, finite-temperature simulations~\cite{Bursill1996, Wang1997, Verstraete2004}, and non-equilibrium systems~\cite{Hieida1998}. Furthermore, it has inspired multigrid DMRG~\cite{Dolfi2012} for systems with multiple length scales and the variational uniform MPS (VUMPS) algorithm~\cite{ZaunerStauber2018} for systems approaching the thermodynamic limit.

In parallel, the conceptual framework of tensor network states has evolved to become recognized as a natural language to describe quantum states of matter~\cite{Evenbly2011, Evenbly2014, Ors2014oct, Ors2014nov, Ors2019, Bauls2023}. The so-called bond dimension parameterizes tensor network states and controls the accuracy, with higher values yielding higher accuracy. In principle, MPS can represent any quantum state provided that the bond dimension $\chi$ is large enough; however, capturing the entanglement structure of higher-dimensional or critical systems requires exponentially large bond dimensions, leading to a rapid increase in computational cost. The reason for the exponential scaling is that MPS only satisfies a one-dimensional boundary law, meaning entanglement saturates to a constant for a fixed $\chi$ as the size of the system increases, making them most effective when representing systems generally corresponding to ground states of local one-dimensional gapped Hamiltonians~\cite{Hastings2006}. This limit explains MPS's challenge with higher-dimensional systems, which often obey higher-dimensional boundary laws, and critical systems, which often obey boundary laws with a multiplicative logarithmic correction. 

Alternative tensor network states have been developed to overcome MPS's inherent limitations, mainly to capture a higher scaling of entanglement entropy effectively. Among these are projected entangled-pair states/operators (PEPS/PEPO)~\cite{Verstraete2004Strongly}, which can be seen as higher-dimensional generalizations of MPS/MPO. Moreover, there are the multi-scale entanglement renormalization ansatz (MERA)~\cite{Vidal2007, Vidal2008} and branching MERA~\cite{Evenbly2014Real, Evenbly2014Class}. Another alternative is the tree tensor network states/operators (TTNS/TTNO)~\cite{Shi2006, Nakatani2013}. Similar to MPS, TTNS, on average, satisfies a one-dimensional boundary law and is also inherently one-dimensional since it lacks cycles. Still, the hierarchical structure offers different benefits to MPS.

However, while these alternatives can effectively capture a higher scaling of entanglement entropy using lower bond dimensions, optimizing them can in total require more work due to increased algorithmic complexity. For reference, the typical computational cost for DMRG scales as $O(\chi_{\text{\tiny MPS}}^3)$~\cite{Stoudenmire2012}. TTNS typically scales as $O(\chi_{\text{\tiny TTNS}}^4)$~\cite{Shi2006, Nakatani2013}. MERA methods, depending on their structure, exhibit computational costs ranging from $O(\chi_{\text{\tiny MERA}}^7)$ to $O(\chi_{\text{\tiny MERA}}^{16})$, with higher costs associated with two-dimensional and branching variants~\cite{Ors2019, Evenbly2009}. Similarly, PEPS will have a typical computational effort that scales as $O(\chi_{\text{\tiny PEPS}}^{10})$ to $O(\chi_{\text{\tiny PEPS}}^{12})$~\cite{Stoudenmire2012}. This stark contrast highlights a trade-off between low bond dimension and high algorithmic complexity.

\section{Methodology}
\label{sec: Methodology}

The survey was performed in two steps. 
First, we searched to identify all packages that implement DMRG, and then filtered out a relevant subset for further investigation (see Section~\ref{sec: DMRG Software Packages} for the results). 
Second, we studied the selected packages to classify their features (see Section~\ref{sec: Comparisons} for the results).

To identify all packages that implement DMRG, we surveyed relevant literature, particularly software release articles, searched online sources and code hosts (e.g., GitHub), and traced citations and related projects. 
The Tensor Network website~\cite{TensorNetwork_web} was a particularly useful starting point.
We believe our coverage of open source packages is relatively complete. 
However, a recurring issue was the lack of transparency in articles regarding the software used for the DMRG calculations. 
This leads us to believe that there could potentially be a large number of packages, especially private ones, that we are not aware of.

To classify the package features, we gathered information from documentation, research articles, and other publicly accessible sources. 
We then approached the developers with questions regarding the features, dependencies, and parallelization strategies of their software to verify and complement the initial information. 
The input from the developers proved invaluable, as the veracity and recency of publicly available information are often lacking. 
Despite our best efforts, errors and omissions may still occur in the presented information. The errors could be due to evolving software features, differing interpretations, or a lack of information.

We maintain updated tables on GitHub~\cite{DMRG-softwareGit} and encourage feedback from package developers and users to help keep the information complete and up-to-date.
\section{Packages Implementing DMRG}
\label{sec: DMRG Software Packages}

The following are brief descriptions of the packages we found and compare in~\Cref{sec: Comparisons}.

\begin{itemize}
    \item \textsc{ALPS} (Algorithms and Libraries for Physics Simulations)~\cite{Albuquerque2007, Bauer2011, ALPS_web, ALPSSourceCode} is a software package aiming to provide standardized components for numerical simulations of condensed matter systems. It has received contributions from numerous researchers worldwide. It contains two separate DMRG packages: \textsc{ALPS DMRG}~\cite{Albuquerque2007, Bauer2011} and \textsc{ALPS MPS}~\cite{Dolfi2014, Bauer2011}.

    \item \textsc{BAGEL} (Brilliantly Advanced General Electronic-structure Library)~\cite{Shiozaki2017, BAGEL_web, BAGELSourceCode} is an electronic structure package maintained by the company QSimulate~\cite{QSimulate_web, QSimulateGitOrg}. In its DMRG implementation, sites are individual molecules that collect multiple orbitals~\cite{Parker2014}. 
    
    \item \textsc{Block2}~\cite{Zhai2023, Zhai2021, Larsson2022, Zhai2022, Block2SourceCode} aims to provide a comprehensive set of DMRG algorithms for use in electronic structure methods and other applications. The package is developed and maintained by the Garnet Chan group at Caltech and the Initiative for Computational Catalysis at the Flatiron Institute. It can interface with other quantum chemistry packages to enable hybrid DMRG methods, including \textsc{PySCF}~\cite{Sun2017, Sun2020}, \textsc{OpenMOLCAS}~\cite{Aquilante2020}, and \textsc{Forte}~\cite{Evangelista2024}.

    \item \textsc{CheMPS2}~\cite{Wouters2014CheMPS2, Wouters2014Density, Wouters2014Communication, Wouters2015, Wouters2016, CheMPS2SourceCode} is a library containing a DMRG implementation for ab initio quantum chemistry primarily developed by Sebastian Wouters. It can interface with quantum chemistry packages that can handle R(O)HF calculations and molecular orbital matrix elements, and this has been done for \textsc{Psi4}~\cite{Smith2020} and \textsc{PySCF}~\cite{Sun2017, Sun2020}.
    
    \item \textsc{ChemTensor}~\cite{ChemTensorSourceCode} is a package for tensor network algorithms centered around chemical systems developed by the Quantum Computing Group at the Technical University of Munich~\cite{QC-TUM_web, QC-TUMGitOrg}.

    \item Chen et al.~\cite{Chen2019, Chen2020} presents a hybrid parallel implementation of DMRG with custom subroutines for vector operations on the GPU for improved performance. More recently, they also presented a novel approach to real-space parallel DMRG \cite{Chen2021}, building upon the work of Stoudenmire and White~\cite{Stoudenmire2013}. 
    
    \item \textsc{Cytnx}~\cite{Wu2025, CytnxSourceCode} is a tensor network library designed for classical and quantum physics simulations. Kai-Hsin Wu is the creator and manages the core team that develops and maintains the package. They also see significant contributions from individuals outside the team. The package has interfaces similar to popular libraries like NumPy~\cite{Harris2020}, SciPy~\cite{Virtanen2020}, and PyTorch~\cite{Ansel2024}. 
    
    \item \textsc{DMRG-Budapest}~\cite{DMRG-Budapest} is a package for general quantum many-body problems. \"{O}rs Legeza leads the Strongly Correlated Systems ``Lend\"{u}let'' Research Group \cite{OrsGroup_web} that develops and maintains the code. It features various novel in-house optimization techniques to improve performance~\cite{Menczer2023, Menczer2024Tensor, Menczer2024Two, Menczer2024Cost, Menczer2024Parallel}.
    
    \item \textsc{DMRG++}~\cite{Alvarez2009, Alvarez2011, Alvarez2012, Alvarez2013, Alvarez_web, DMRG++SourceCode} implements the DMRG algorithm with an emphasis on generic programming and minimal software dependencies. The main developer is Gonzalo Alvarez. Additionally, the package has a plug-in to extend its capabilities to GPUs~\cite{Elwasif2018, dmrgppPluginScSourceCode}.
    
    \item \textsc{DMRGPy}~\cite{DMRGPySourceCode} is a library to simulate quasi-one-dimensional spin chains and fermionic systems, mainly developed by Jose Lado. The library relies on \textsc{ITensor}~\cite{Fishman2022, ITensor_web, ITensorGitOrg} (either the Julia or the \CC{} version). 

    \item \textsc{FOCUS}~\cite{Xiang2024} implements an ab initio DMRG algorithm. Zhendong Li originally introduced the package with a pilot implementation of a DMRG-like sweep algorithm for variational optimization of comb tensor network states~\cite{Li2021}. It supports both non-relativistic and relativistic Hamiltonians, utilizing time-reversal symmetry in the latter case to reduce computational costs~\cite{Li2023}.

    \item Hong et al.~\cite{Hong2022} presents a DMRG implementation using various optimization strategies, specifically targeting GPUs with tensor cores.

    \item \textsc{ITensor}~\cite{Fishman2022, ITensor_web, ITensorGitOrg} is a package for programming tensor network calculations, allowing users to focus on the connectivity of a tensor network without manually tracking indices. Matt Fishman and Miles Stoudenmire are the primary developers and receive contributions from collaborators. \textsc{ITensor} was initially implemented in \CC{} but later fully ported to Julia, with most new features first being developed there. The \CC{} version~\cite{ITensorSourceCode} includes DMRG in the main library, while the Julia version provides DMRG through the \textsc{ITensorMPS.jl}~\cite{ITensorMPS.jlSourceCode} extension of \textsc{ITensors.jl}~\cite{ITensors.jlSourceCode}.

    \item \textsc{Kylin}~\cite{Xie2023, Kylin_web} is an ab initio quantum chemistry software package for estimating the electronic structures of molecular systems. The package is developed by the Ma Group~\cite{MaResearchGroup_web} at Shandong University, and it is specially designed for calculations with large active spaces.

    \item \textsc{MOLMPS}~\cite{Brabec2020} is a parallel implementation of DMRG for quantum chemistry. The Libor Veis Research Group~\cite{LiborVeisResearchGroup_web} at the Heyrovský Institute~\cite{HeyrovskýInstitute_web} develops and maintains the code. The parallel scheme is based on an in-house MPI global memory library and supports various hybrid electronic structure methods.

    \item \textsc{MPSKit.jl}~\cite{van_damme_2024_14395415, MPSKit.jlSourceCode} contains tensor network algorithms for one-dimensional quantum and two-dimensional statistical mechanics problems. It is developed under the QuantumKitHub organization~\cite{QuantumKitHubGitOrg}, where many related software packages focusing on quantum many-body physics through tensor networks are publicly available, with contributions mainly from the Quantum Group at Ghent University~\cite{QuantumGroupGhent_web}. It mainly builds upon \textsc{TensorKit.jl}~\cite{jutho_2025_14632990, TensorKit.jlSourceCode}, which provides functionality for generic symmetries, and its capabilities can be extended with packages like \textsc{SUNRepresentations.jl}~\cite{SUNRepresentations.jlSourceCode} and \textsc{MPSKitModels.jl}~\cite{MPSKitModels.jlSourceCode}.

    \item \textsc{MPToolkit} (Matrix Product Toolkit)~\cite{MPToolkitGitOrg, MPToolkitSourceCode} is a package for creating and manipulating MPS. Ian McCulloch started the project around 2002, which was initially envisioned as a ``next-generation'' DMRG code that integrates non-abelian symmetries and emphasizes a flexible, general approach to constructing Hamiltonian operators and measuring observables~\cite{McCulloch2002}.

    \item \textsc{OSMPS} (Open Source MPS)~\cite{Wall2012, Jaschke2018apr, Jaschke2018oct, Jaschke2018nov, OSMPSSourceCode} is a collection of tensor network algorithms for simulating entangled one-dimensional many-body quantum systems. The package is developed by Michael L. Wall, Daniel Jaschke, Matthew Jones, and the Carr Theoretical Physics Research Group~\cite{CarrResearchGroup_web}. The initial focus was to offer a broad set of methods for quantum simulators based on atomic, molecular, and optical physics architectures. 

    \item \textsc{PyTeNet}~\cite{Mendl2018, PyTeNetSourceCode} implements quantum tensor network operations and simulations structured around MPS and MPO classes. The Quantum Computing Group at the Technical University of Munich~\cite{QC-TUM_web, QC-TUMGitOrg} develops the package. It acts as a facilitator of algorithmic experimentation. 

    \item \textsc{pyTTN}~\cite{Lindoy2025, pyTTNSourceCode} is a package for working with generic TTNS to compute dynamical properties of quantum systems, primarily developed by Lachlan Lindoy. It also provides functionality for merging many physical modes into a single site, which can be useful when handling weakly correlated subsystems with a strong correlation between the degrees of freedom in each subsystem. 
    
    \item \textsc{QCMaquis}~\cite{Keller2015, Knecht2016, QCMaquis_web, QCMaquisSourceCode} is a SCINE module~\cite{SCINE_web, SCINEGitOrg} that builds upon the \textsc{ALPS MPS}~\cite{Dolfi2014, Bauer2011} code and implements various DMRG-based algorithms. The Reiher Research  Group at ETH Zurich~\cite{ReiherResearchGroup_web, SCINEGitOrg} develops the package. Key features include vibrational~\cite{Baiardi2017, Baiardi2019mar}, time-dependent~\cite{Baiardi2019may, Baiardi2021}, and nuclear-electron~\cite{Muolo2020, Feldmann2022} DMRG, and it accommodates non-relativistic and relativistic electronic structure calculations~\cite{Battaglia2018}.
    
    \item \textsc{QSpace}~\cite{Weichselbaum2012, Weichselbaum2020, Weichselbaum2024, QSpaceSourceCode} is a tensor library designed as a bottom-up approach for non-abelian symmetries, starting from the defining representation and the respective Lie algebra. It has a long-standing history, with Andreas Weichselbaum as the primary developer. A distinctive feature is its versatility in operations across all symmetries, permitting arbitrary combinations. 
    
    \item \textsc{Quantum TEA} (Quantum Tensor network Emulator Applications)~\cite{bacilieri_2024_13383350, QuantumTEA_web, QuantumTEASourceCode} is a set of tensor network packages for quantum simulation, circuit emulation, and machine learning applications. The Quantum Information and Matter group at the University of Padova~\cite{QuantumInformationandMattergroup_web} develops the package. Supplementary libraries extend their capabilities with diverse tensor backends to enable computations on CPUs, GPUs, and TPUs~\cite{jaschke2024benchmarkingquantumredtea}. 

    \item \textsc{quimb}~\cite{Gray2018, quimbSourceCode} is a library for quantum information and many-body calculations, focusing primarily on tensor networks, developed mainly by Johnnie Gray. While its DMRG routine does not currently support features commonly found in other packages, \textsc{quimb} is compatible with complementary libraries like \textsc{cotengra}~\cite{Gray2021, cotengraGit}, \textsc{autoray}~\cite{autorayGit}, and \textsc{symmray}~\cite{symmrayGit} to support efficient tensor network contraction, and various backend array libraries supporting block-sparse, abelian-symmetric, and fermionic representations.
    
    \item \textsc{Renormalizer}~\cite{Ren2018, Li2020, Jiang2020, Ren2022, RenormalizerSourceCode} is a tensor network package with a focus on electron-phonon quantum dynamics developed by the Zhigang Shuai group~\cite{shuaigroup_web, shuaigroupGitOrg}. It implements an original MPO/TTNO construction algorithm that leverages bipartite graph theory to automate and optimize the selection of normal and complementary operators for custom Hamiltonians in the sum-of-products form~\cite{Ren2020, Li2024}, a feature later adopted by several other software packages.
    
    \item \textsc{SeeMPS2}~\cite{GarcaRipoll2021, GarcaRipoll2023, SeeMPS2SourceCode} is the second iteration of the SElf-Explaining Matrix-Product-State library~\cite{SeeMPSSourceCode}. Juan José García Ripoll and Paula García Molina in the Quantum Information and Foundations Group at the Institute of Fundamental Physics, in the Spanish Research Council (CSIC)~\cite{QUINFOGroup_web} develop the package. It aims to enable rapid prototyping and testing of MPS and DMRG-inspired algorithms. 

    \item \textsc{SUNDMRG.jl}~\cite{SUNDMRG.jlSourceCode} is a DMRG implementation specializing in a full $\mathrm{SU}(n)$ symmetry and is developed by Masahiko G. Yamada, James R. Garrison, and Ryan V. Mishmash.

    \item \textsc{SymMPS}~\cite{SymMPS_web, SymMPSSourceCode} builds upon the Scientific Parallel Algorithms Library (SciPAL)~\cite{Kramer2015} and is developed by Sebastian Paeckel and Thomas Kölner. It features an original construction scheme for MPO representations of arbitrary $\mathrm{U}(1)$-symmetric operators whenever there is an expression of the local structure in terms of a finite-state machine~\cite{Paeckel2017}.
    
    \item \textsc{SyTen}~\cite{hubig17:_symmet_protec_tensor_networ, SyTen_web} aims to be a tensor network toolkit with standard MPS, binary TTNS, and infinite PEPS utilities. Claudius Hubig created the package, and Sebastian Paeckel currently maintains it. It has also benefited from significant contributions from many others, primarily members of the group of Uli Schollwöck at LMU Munich~\cite{SchollwöckGroup_web}.
    
    \item \textsc{TeNPy} (Tensor Network Python)~\cite{Hauschild2018, Hauschild2024, TeNPySourceCode} is a library for simulating strongly correlated quantum systems with tensor networks. Johannes Hauschild and Jakob Unfried are the primary developers, and many others have made significant contributions. Their stated philosophy is to balance readability and usability for newcomers while providing powerful algorithms for experts.
    
    \item \textsc{tensor-tools}~\cite{tensor-toolsConf, tensor-toolsSourceCode} is a tensor network library that builds upon the Cyclops Tensor Framework (CTF)~\cite{Solomonik2014}. Ryan Levy developed the parallel version, which builds upon the serial version created by Xiongjie Yu, both of which Bryan Clark guided as they were part of the Clark Research Group~\cite{ClarkResearchGroup_web}. The package utilizes multiple \textsc{ITensor}~\cite{Fishman2022, ITensor_web, ITensorGitOrg} systems and design concepts to improve interoperability between the two.

    \item \textsc{TensorTrack}~\cite{devos_2024_10641042, TensorTrackSourceCode} is a package implementing various elementary algorithms that arise in the context of tensor networks and is developed by the Quantum Group at Ghent University~\cite{QuantumGroupGhent_web}. The package supports generic symmetries, including symmetry groups with multiplicities, which can have either bosonic or fermionic braiding rules.
    
    \item \textsc{UltraDMRG}~\cite{UltraDMRGSourceCode} is a library for performing large-scale calculations using one-dimensional tensor network algorithms developed by Hao-Xin Wang. It uses \textsc{TensorToolkit}~\cite{TensorToolkitGit}, also developed by Hao-Xin Wang, for tensor operations and is specifically designed to tackle two-dimensional strongly correlated electron systems. 
    
    \item \textsc{xDMRG++}~\cite{xDMRG++SourceCode} is a package that includes various MPS algorithms for studying one-dimensional quantum spin chains. David Aceituno at KTH Condensed Matter Physics group developed the package.

    \item \textsc{YASTN}~\cite{Rams2025, YASTNSourceCode} (Yet Another Symmetric Tensor Network) is a package for differentiable linear algebra with block-sparse tensors, supporting abelian symmetries. Marek Rams and Juraj Hasik are the primary developers. The package mainly focuses on PEPS but also supports a range of MPS algorithms.
\end{itemize}

We also identified a few additional packages, which we decided to exclude from \Cref{sec: Comparisons} for various reasons. We mention them here for completeness. The following appear to no longer be actively maintained and/or used: \textsc{DMRG.x}~\cite{Vance2017DMRGx, DMRG.xGit}, \textsc{mps}~\cite{mpsGit}, \textsc{MPS++}~\cite{MPS++Git}, \textsc{MPSTools}~\cite{MPSToolsGit}, \textsc{ORZ}~\cite{Kurashige2013}, \textsc{parallelDMRG}~\cite{Stoudenmire2013, parallelDMRGGit}, \textsc{PyDMRG}~\cite{PyDMRGGit}, \textsc{PyTeN}~\cite{PyTeNGit}, \textsc{QC-MPO-DMRG}~\cite{qcmpodmrgGit}, \textsc{REGO}~\cite{Kurashige2009}, \textsc{Snake DMRG} \cite{Guo2009, snake-dmrgGit}, \textsc{TensorNetwork}~\cite{roberts2019tensornetworklibraryphysicsmachine, TensorNetworkGit}, \textsc{TNT} \cite{AlAssam2017}, and \textsc{ZMPO-DMRG}~\cite{zmpo_dmrgGit}. Next are those that have been superseded by other packages: \textsc{Block} \cite{blockGit} and \textsc{StackBlock} \cite{stackblockGit} are precursors to \textsc{Block2}~\cite{Zhai2023, Block2SourceCode}, \textsc{GraceQ/MPS2}~\cite{GraceQMPS2Git} is a precursor to \textsc{UltraDMRG}~\cite{UltraDMRGSourceCode}, and \textsc{Uni10}~\cite{Kao2015, uni10Git} is a precursor to \textsc{Cytnx}~\cite{Wu2025, CytnxSourceCode}. Finally, \textsc{Simple DMRG}~\cite{james_r_garrison_2017_1068359}, \textsc{sophisticated-dmrg}~\cite{james_r_garrison_2021_4651419}, Glen Evenbly's example codes~\cite{Tensors.net_web}, and G. Catarina's implementation~\cite{Catarina2023} were designed primarily as tutorials rather than scalable research tools and were therefore excluded.

\section{Feature Comparisons}
\label{sec: Comparisons}

In this section, we present and compare multiple aspects of the package features.
We emphasize that our classification focuses on the functionality of these packages within the context of DMRG rather than serving as a qualitative assessment of their effectiveness. The presence of a feature does not imply anything about its efficiency or performance. Hence, a package with more features is not necessarily more versatile or higher-performing than others. Additionally, many features, such as out-of-core capabilities, subspace expansion techniques, permitted boundary conditions, and lattice geometries, were not investigated. We intentionally focused on general features to provide a broad and informative classification.

\subsection{High-Level Overview}
\label{ssec: High-Level Overview}

We present some high-level aspects of each package in \Cref{tab: main}. We highlight the implementation language and scheme, support for symmetries, and HPC capabilities. The columns in \Cref{tab: main} are as follows:
\begin{itemize}
    \item \emph{ID:} A unique identifier for cross-referencing purposes.

    \item \emph{Name:} The official name of the package; if there is one.

    \item \emph{Language:} The programming languages used for implementation and interfaces. Languages marked with \textsuperscript{i}-superscripts are those mainly used as interfaces.
    
    \item \emph{Open Source Software (OSS):} Indicates whether the package is open source and available for download.
    
    \item \emph{Implementation Formalism (Gen.):} Specifies which view the DMRG implementation is based on: 1\textsuperscript{st} for the traditional RG perspective using renormalized operators and 2\textsuperscript{nd} for the tensor network perspective using MPO/MPS. Additionally, \textsuperscript{$\ddagger$}-superscripts mark those who support both views, and \textsuperscript{$\dagger$}-superscripts mark those who support the usage of TTNO/TTNS.

    \item \emph{Symmetry Support (Sym.):} Indicates whether the package is symmetry-adapted to support abelian (A) and/or non-abelian (NA) symmetries or not.

    \item \emph{High-Performance Computing (HPC):} Indicates support for HPC platforms, categorized as shared-memory parallelism (SM), distributed-memory parallelism (DM), and single- (S) or multi- (M) GPU acceleration (GPU). 
\end{itemize}

\begin{table}[htbp]
    \centering
    \caption{High-level overview of the packages included in the survey. Note that support for multiple features does not imply that they can be utilized simultaneously.}
    \begin{tabular}{lllclccccc}
\toprule
\multirow{2}{*}{{\bf ID}} & \multirow{2}{*}{{\bf Name}} & \multirow{2}{*}{{\bf Language}} & \multirow{2}{*}{{\bf OSS}} & \multirow{2}{*}{{\bf Gen.}} & \multicolumn{2}{c}{{\bf Sym.}} & \multicolumn{3}{c}{{\bf HPC}} \\
\cmidrule(lr){6-7}\cmidrule(lr){8-10}
 &  &  &  &  & {\bf A} & {\bf NA} & {\bf SM} & {\bf DM} & {\bf GPU} \\
\midrule
1 & ALPS DMRG & \CC, Python\textsuperscript{i} & \cite{ALPSSourceCode} & 1\textsuperscript{st} & \checkmark & - & \checkmark & - & - \\ \hline
2 & ALPS MPS & \CC, Python\textsuperscript{i} & \cite{ALPSSourceCode} & 2\textsuperscript{nd} & \checkmark & - & \checkmark & - & - \\ \hline
3 & BAGEL & \CC & \cite{BAGELSourceCode} & 1\textsuperscript{st} & \checkmark & - & \checkmark & \checkmark & - \\ \hline
4 & Block2 & \CC, Python\textsuperscript{i} & \cite{Block2SourceCode} & 2\textsuperscript{nd}\textsuperscript{$\ddagger$} & \checkmark & \checkmark & \checkmark & \checkmark & - \\ \hline
5 & CheMPS2 & \CC, Python\textsuperscript{i} & \cite{CheMPS2SourceCode} & 1\textsuperscript{st} & \checkmark & \checkmark & \checkmark & \checkmark & - \\ \hline
6 & ChemTensor & C, Python\textsuperscript{i} & \cite{ChemTensorSourceCode} & 2\textsuperscript{nd}\textsuperscript{$\dagger$} & \checkmark & \checkmark & \checkmark & - & - \\ \hline
7 & Chen et al. & \CC & - & 1\textsuperscript{st} & \checkmark & - & \checkmark & - & S \\ \hline
8 & Cytnx & \CC, Python\textsuperscript{i} & \cite{CytnxSourceCode} & 2\textsuperscript{nd}\textsuperscript{$\dagger$} & \checkmark & - & \checkmark & - & S \\ \hline
9 & DMRG-Budapest & \CC, MATLAB\textsuperscript{i} & - & 2\textsuperscript{nd}\textsuperscript{$\ddagger$} & \checkmark & \checkmark & \checkmark & \checkmark & M \\ \hline
10 & DMRG++ & \CC & \cite{DMRG++SourceCode} & 1\textsuperscript{st} & \checkmark & - & \checkmark & - & S \\ \hline
11 & DMRGPy & Python & \cite{DMRGPySourceCode} & 2\textsuperscript{nd} & - & - & \checkmark & - & - \\ \hline
12 & FOCUS & \CC & - & 1\textsuperscript{st} & \checkmark & \checkmark & \checkmark & \checkmark & M \\ \hline
13 & Hong et al. & \CC & - & 2\textsuperscript{nd} & - & - & - & - & S \\ \hline
14 & ITensor & \CC & \cite{ITensorSourceCode} & 2\textsuperscript{nd} & \checkmark & - & \checkmark & - & - \\ \hline
15 & ITensorMPS.jl & Julia & \cite{ITensorMPS.jlSourceCode} & 2\textsuperscript{nd} & \checkmark & - & \checkmark & - & S \\ \hline
16 & Kylin & \CC & - & 2\textsuperscript{nd} & \checkmark & \checkmark & \checkmark & - & - \\ \hline
17 & MOLMPS & \CC & - & 1\textsuperscript{st} & \checkmark & - & \checkmark & \checkmark & S \\ \hline
18 & MPSKit.jl & Julia & \cite{MPSKit.jlSourceCode} & 2\textsuperscript{nd} & \checkmark & \checkmark & \checkmark & - & - \\ \hline
19 & MPToolkit & \CC & \cite{MPToolkitSourceCode} & 2\textsuperscript{nd} & \checkmark & \checkmark & \checkmark & - & - \\ \hline
20 & OSMPS & Fortran, Python\textsuperscript{i} & \cite{OSMPSSourceCode} & 2\textsuperscript{nd} & \checkmark & - & \checkmark & \checkmark & - \\ \hline
21 & PyTeNet & Python & \cite{PyTeNetSourceCode} & 2\textsuperscript{nd} & \checkmark & - & \checkmark & - & - \\ \hline
22 & pyTTN & \CC, Python\textsuperscript{i} & \cite{pyTTNSourceCode} & 2\textsuperscript{nd}\textsuperscript{$\dagger$} & - & - & \checkmark & - & - \\ \hline
23 & QCMaquis & \CC & \cite{QCMaquisSourceCode} & 2\textsuperscript{nd} & \checkmark & \checkmark & \checkmark & - & - \\ \hline
24 & QSpace & \CC, MATLAB\textsuperscript{i} & \cite{QSpaceSourceCode} & 2\textsuperscript{nd}\textsuperscript{$\ddagger$} & \checkmark & \checkmark & \checkmark & - & - \\ \hline
25 & Quantum TEA & Fortran, Python\textsuperscript{i} & \cite{QuantumTEASourceCode} & 2\textsuperscript{nd}\textsuperscript{$\dagger$} & \checkmark & - & \checkmark & - & S \\ \hline
26 & quimb & Python & \cite{quimbSourceCode} & 2\textsuperscript{nd} & - & - & \checkmark & - & - \\ \hline
27 & Renormalizer & Python & \cite{RenormalizerSourceCode} & 2\textsuperscript{nd}\textsuperscript{$\dagger$} & \checkmark & - & \checkmark & - & S \\ \hline
28 & SeeMPS2 & Python & \cite{SeeMPS2SourceCode} & 2\textsuperscript{nd} & - & - & \checkmark & - & - \\ \hline
29 & SUNDMRG.jl & Julia & \cite{SUNDMRG.jlSourceCode} & 1\textsuperscript{st} & - & \checkmark & \checkmark & \checkmark & S \\ \hline
30 & SymMPS & \CC & \cite{SymMPSSourceCode} & 2\textsuperscript{nd} & \checkmark & - & \checkmark & - & - \\ \hline
31 & SyTen & \CC, Python\textsuperscript{i} & - & 2\textsuperscript{nd} & \checkmark & \checkmark & \checkmark & \checkmark & S \\ \hline
32 & TeNPy & Python & \cite{TeNPySourceCode} & 2\textsuperscript{nd} & \checkmark & - & \checkmark & \checkmark & - \\ \hline
33 & tensor-tools & \CC & \cite{tensor-toolsSourceCode} & 2\textsuperscript{nd} & \checkmark & - & \checkmark & \checkmark & - \\ \hline
34 & TensorTrack & MATLAB & \cite{TensorTrackSourceCode} & 2\textsuperscript{nd} & \checkmark & \checkmark & \checkmark & - & - \\ \hline
35 & UltraDMRG & \CC & \cite{UltraDMRGSourceCode} & 2\textsuperscript{nd} & \checkmark & - & \checkmark & \checkmark & S \\ \hline
36 & xDMRG++ & \CC & \cite{xDMRG++SourceCode} & 2\textsuperscript{nd} & - & - & \checkmark & - & - \\ \hline
37 & YASTN & Python & \cite{YASTNSourceCode} & 2\textsuperscript{nd} & \checkmark & - & \checkmark & - & S \\ 
\bottomrule
\end{tabular}

    \label{tab: main}
\end{table}

To provide further clarity regarding \Cref{tab: main}, firstly, most developers of packages listed as not open source are willing to share their work in collaborations on specific projects or may offer it for free to people in academia.

Secondly, DMRG expressed with tensor network operators/states is sometimes called a second-generation formalism and allows for more flexibility in finding the optimal state
with a broader range of applications~\cite{Keller2015}, hence the distinction. However, it is essential to note that both can be equally efficient, and to a certain extent, the difference between them is mainly semantic in the context of DMRG. In an expectation value optimization, the first-generation renormalized operators naturally correspond to partial traces of the second-generation MPO. In DMRG, these are the only ones needed; the explicit MPO itself never needs to appear. However, for calculations of higher moments, for example, an MPO is demonstrably more accurate as it exactly represents the Hamiltonian, unlike the renormalized operators~\cite{Schollwck2011, Chan2016}.

Thirdly, we classify packages based on whether they support symmetry in any form, without distinguishing between those limited to specific models and those that allow arbitrary symmetry combinations. \Cref{tab: symmetry} provides more fine-grained symmetry distinctions.

Finally, similar to the symmetries, we classify packages based on whether they support HPC platforms in any form, regardless of their use. \Cref{tab: HPC strategies} provides a more detailed examination of the different strategies and optimization techniques. 

We conclude that most packages are implemented in \CC{} (often with a Python interface) and some in Python. 
A few are implemented in Julia, Fortran, or MATLAB.
Most support some forms of symmetries and shared-memory parallelism. 
A minority of the packages support distributed-memory and/or GPU acceleration.
\subsection{Dependencies}

All surveyed packages have external software dependencies. 
Some are mandatory while others are optional. 
Some packages are used in tandem with other packages that provide additional features, even though one is not technically dependent on the other. 
For example, the package \textsc{MPSKit.jl} can be used together with the package \textsc{MPSKitModels.jl}, which provides concrete operators and models. 
For the purposes of this section, we also consider these relationships as dependencies. 

\begin{figure}[hbtp]
    \centering
    \includegraphics[width=\textwidth]{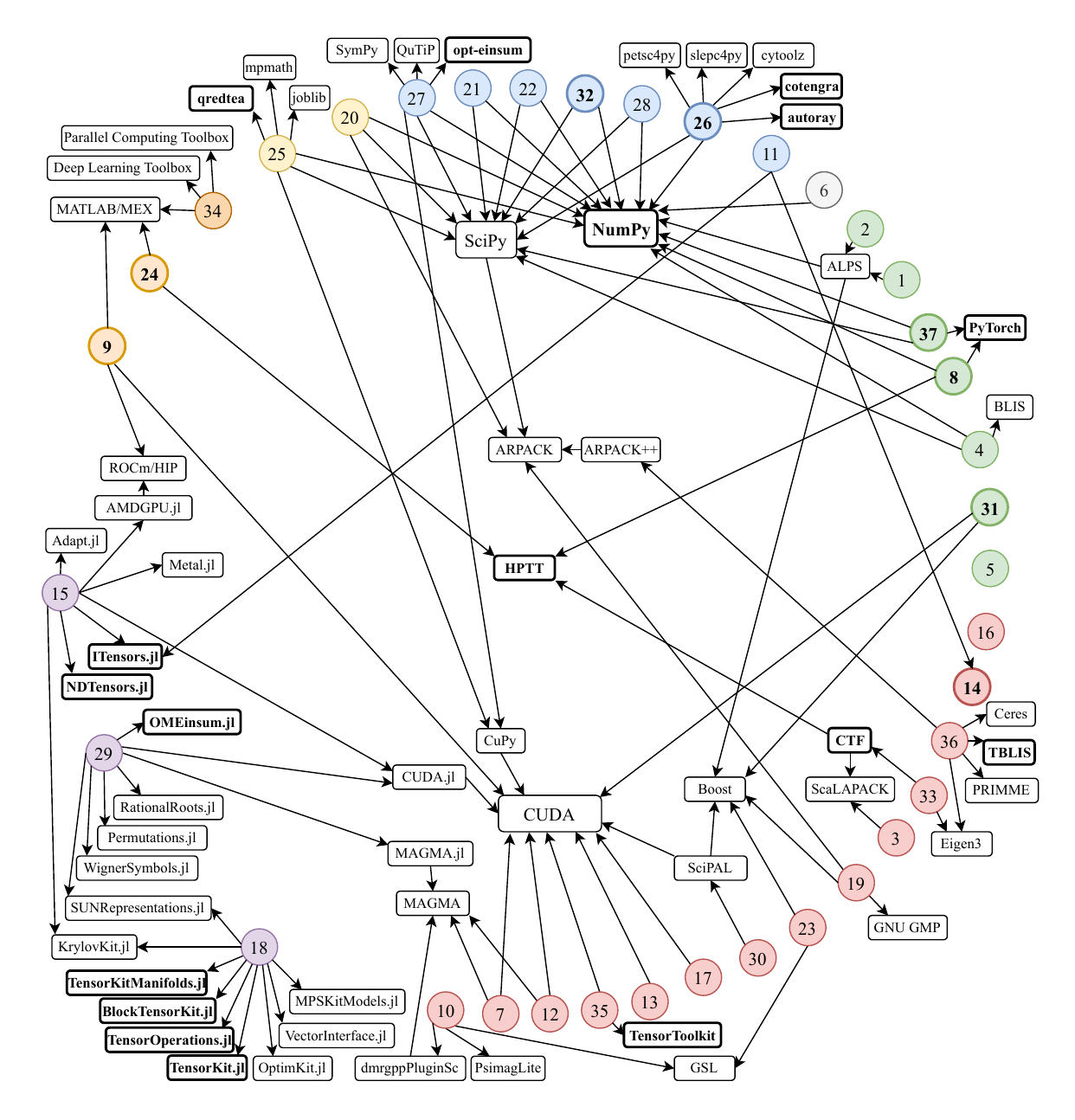}
    \caption{Graph showing the required and optional dependencies, and compatible software relevant to DMRG calculations for the different packages. The circle numbers correspond to those in \Cref{tab: main}. The circles are color-coded based on the \emph{Language}-column in \Cref{tab: main}. We have highlighted tensor-related libraries with a bold font and thicker outline.
    }
    \label{fig: dependency graph}
\end{figure}

We present the dependencies of each package in \Cref{fig: dependency graph}.
Only those dependencies most relevant for DMRG calculations are included. 
For example, dependencies related to parsing and serialization were excluded.
Thus, despite their prevalence, dependencies such as HDF5 are omitted. 
Similarly, Boost is not marked as a dependency when the documentation states that it is only used for parsing and serialization.
We do not distinguish between required and optional dependencies. 
We also exclude edges between third-party libraries as long as this does not affect the set of dependencies reachable from any package.

Since all packages depend on some implementation of BLAS and LAPACK, we excluded them to reduce clutter. 
Furthermore, we excluded dependencies associated with MPI and threading (e.g., OpenMP). 
These dependencies are implicitly captured in \Cref{tab: main,tab: HPC strategies}. 
We grouped CUDA-related software into a single node, covering bespoke CUDA kernels, tensor and other numerical libraries (cuTT, cuTENSOR, cuQUANTUM, cuBLAS, cuSOLVER), communication (NCCL), and core components (CUDA Runtime, CUDA Toolkit). 

We conclude from this analysis that the packages are mostly independent, with little common functionality factored out into third-party libraries.
The main exceptions are BLAS/LAPACK, which all packages rely on directly or indirectly, and the standard parallel support frameworks (e.g., OpenMP, MPI, CUDA).
The Python-based packages commonly use SciPy and NumPy.
Some libraries provide higher-level tensor operations, but they are mostly used by one or two packages. 

\subsection{Parallelism and Mixed-Precision}

We now expand on the HPC aspects introduced in \Cref{ssec: High-Level Overview} to describe parallelization strategies and other optimization techniques. We closely follow the classification summary and five-level hierarchy of parallelism first introduced by Zhai and Chan~\cite{Zhai2021} and then further discussed by Tian and Ma~\cite{Tian2023}. \Cref{tab: HPC strategies} presents the following:
\begin{itemize}
    \item \emph{Parallel strategies (Parallelism):} Indicates support for the following types of parallelism:
    \begin{itemize}
        \item \emph{Parallelism within matrix operations (i):} The most fine-grained and lowest-level source of parallelism is the data parallelism found primarily within the matrix--matrix and matrix--vector multiplications that underpin the tensor algebra in the DMRG algorithm.

        \item \emph{Parallelism over symmetry sectors (ii):} The block-sparse tensor representations in symmetry-adapted DMRG implementations enable parallelism across symmetry sectors. The calculations on different blocks are independent, allowing them to run simultaneously~\cite{Kurashige2009, Solomonik2014}.
        
        \item \emph{Parallelism over normal and complementary operators (iii):} In ab initio DMRG, the left-right decomposition of the Hamiltonian is expressed as a summation of products of normal and complementary operators. These operators are distributed over threads and processors, processing the corresponding calculations concurrently~\cite{Chan2004, Wouters2014Density}.
        
        \item \emph{Parallelism over a sum of sub-Hamiltonians (iv):} The Hamiltonian is rewritten as a sum of sub-Hamiltonians such that each term can be manipulated independently~\cite{Chan2016}. This coarse-grained, high-level parallelism is easily expressed in the MPO formalism and adds little communication overhead.
        
        \item \emph{Parallelism over sites (v):} The DMRG sequentially sweeps back and forth along a path. The path can be partitioned into sections that can be processed simultaneously at the expense of extra communication at the section boundaries~\cite{Stoudenmire2013, Chen2021}. This coarse-grained parallelism is especially useful for large systems with many sites.
        
    \end{itemize}
    
    \item \emph{Mixed-precision (MP):} Indicates support for the recent mixed-precision optimization technique introduced by Tian et al.~\cite{Tian2022}, where the initial sweeps are accelerated by using reduced floating-point precision. Later sweeps restore full precision by switching to full floating-point precision.
\end{itemize}

\begin{table}[hbtp]
    \centering
    \caption{Support for parallelism strategies and mixed-precision optimization techniques. Note that support for multiple features does not imply that they can be utilized simultaneously.}
    \begin{tabular}{llccccccccc}
\toprule
\multirow{2}{*}{{\bf ID}} & \multirow{2}{*}{{\bf Name}} & \multicolumn{3}{c}{{\bf HPC}} & \multicolumn{5}{c}{{\bf Parallelism}} & \multirow{2}{*}{{\bf MP}} \\
\cmidrule(lr){3-5}\cmidrule(lr){6-10}
 &  & {\bf SM} & {\bf DM} & {\bf GPU} & {\bf i} & {\bf ii} & {\bf iii} & {\bf iv} & {\bf v} &  \\
\midrule
1 & ALPS DMRG & \checkmark & - & - & \checkmark & - & - & - & - & - \\ \hline
2 & ALPS MPS & \checkmark & - & - & \checkmark & \checkmark & - & - & - & - \\ \hline
3 & BAGEL & \checkmark & \checkmark & - & \checkmark & - & - & - & - & - \\ \hline
4 & Block2 & \checkmark & \checkmark & - & \checkmark & \checkmark & \checkmark & \checkmark & \checkmark & \checkmark \\ \hline
5 & CheMPS2 & \checkmark & \checkmark & - & - & \checkmark & \checkmark & - & - & - \\ \hline
6 & ChemTensor & \checkmark & - & - & - & \checkmark & - & \checkmark & - & - \\ \hline
7 & Chen et al. & \checkmark & - & S & \checkmark & \checkmark & - & - & \checkmark & - \\ \hline
8 & Cytnx & \checkmark & - & S & \checkmark & \checkmark & \checkmark & - & - & - \\ \hline
9 & DMRG-Budapest & \checkmark & \checkmark & M & \checkmark & \checkmark & \checkmark & \checkmark & - & \checkmark \\ \hline
10 & DMRG++ & \checkmark & - & S & \checkmark & \checkmark & - & - & - & - \\ \hline
11 & DMRGPy & \checkmark & - & - & \checkmark & - & - & - & - & - \\ \hline
12 & FOCUS & \checkmark & \checkmark & M & \checkmark & \checkmark & \checkmark & - & - & - \\ \hline
13 & Hong et al. & - & - & S & \checkmark & - & - & - & - & - \\ \hline
14 & ITensor & \checkmark & - & - & \checkmark & \checkmark & - & - & - & - \\ \hline
15 & ITensorMPS.jl & \checkmark & - & S & \checkmark & \checkmark & - & - & - & \checkmark \\ \hline
16 & Kylin & \checkmark & - & - & - & \checkmark & \checkmark & - & - & \checkmark \\ \hline
17 & MOLMPS & \checkmark & \checkmark & S & \checkmark & \checkmark & \checkmark & - & - & - \\ \hline
18 & MPSKit.jl & \checkmark & - & - & \checkmark & \checkmark & - & \checkmark & - & - \\ \hline
19 & MPToolkit & \checkmark & - & - & \checkmark & \checkmark & - & - & - & - \\ \hline
20 & OSMPS & \checkmark & \checkmark & - & \checkmark & \checkmark & - & - & - & - \\ \hline
21 & PyTeNet & \checkmark & - & - & \checkmark & - & - & - & - & - \\ \hline
22 & pyTTN & \checkmark & - & - & \checkmark & - & - & \checkmark & - & - \\ \hline
23 & QCMaquis & \checkmark & - & - & - & \checkmark & \checkmark & \checkmark & - & - \\ \hline
24 & QSpace & \checkmark & - & - & \checkmark & \checkmark & - & - & - & \checkmark \\ \hline
25 & Quantum TEA & \checkmark & - & S & \checkmark & - & - & - & - & \checkmark \\ \hline
26 & quimb & \checkmark & - & - & \checkmark & - & - & - & - & - \\ \hline
27 & Renormalizer & \checkmark & - & S & \checkmark & - & - & - & - & - \\ \hline
28 & SeeMPS2 & \checkmark & - & - & \checkmark & - & - & - & - & - \\ \hline
29 & SUNDMRG.jl & \checkmark & \checkmark & S & \checkmark & \checkmark & - & - & - & - \\ \hline
30 & SymMPS & \checkmark & - & - & \checkmark & \checkmark & - & - & - & - \\ \hline
31 & SyTen & \checkmark & \checkmark & S & \checkmark & \checkmark & - & \checkmark & \checkmark & - \\ \hline
32 & TeNPy & \checkmark & \checkmark & - & \checkmark & \checkmark & - & \checkmark & - & - \\ \hline
33 & tensor-tools & \checkmark & \checkmark & - & \checkmark & - & - & - & - & - \\ \hline
34 & TensorTrack & \checkmark & - & - & \checkmark & - & - & - & - & - \\ \hline
35 & UltraDMRG & \checkmark & \checkmark & S & \checkmark & \checkmark & - & \checkmark & - & - \\ \hline
36 & xDMRG++ & \checkmark & - & - & \checkmark & - & - & - & - & - \\ \hline
37 & YASTN & \checkmark & - & S & \checkmark & - & - & - & - & - \\ 
\bottomrule
\end{tabular}

    \label{tab: HPC strategies}
\end{table}

For a more comprehensive understanding of parallelism within matrix operations, the DMRG algorithm utilizes tensor data structures and related tensor algebra, which includes contraction, diagonalization, and decomposition. The sparse matrix--vector multiplication is the dominant operation of the diagonalization done by the iterative eigensolver and can be decomposed into several dense matrix--matrix multiplications, which can be computed in parallel. Since all packages depend on BLAS/LAPACK in some way, the parallelization of the dense matrix--matrix multiplications can often be trivially implemented by simply linking to a multi-threaded or GPU-accelerated version~\cite{Hager2004, Solomonik2014}.

Despite the simplicity of implementing parallel matrix--matrix multiplications, some packages reported not supporting strategy \emph{i}. By taking a closer look, we see that all of them are related to quantum chemistry. It is unlikely that they do not actually support it; more likely is that they do not prioritize it. They likely prioritize strategy \emph{iii} instead, as it is often regarded as the largest source of parallelism for typical ab initio problems in quantum chemistry~\cite{Zhai2021, Tian2023}. Another reason for this prioritization could be because the matrices that form the operands for matrix--matrix multiplication calls are often relatively small in DMRG, leading to non-negligible parallelization overhead~\cite{Hager2004, Li2020}. Additionally, strategy \emph{iii} is closely connected to the parallelization of the sparse matrix--vector multiplication, so it could be that strategy \emph{i} was thought of as only being parallel matrix--matrix multiplications by the developers when asked. 

Regarding strategy \emph{v}, Chen et al. \cite{Chen2021} recently presented an improvement of the original scheme by Stoudenmire et al. \cite{Stoudenmire2013}, allowing section boundaries to be adaptively distributed. However, we do not distinguish which specific scheme is implemented in \Cref{tab: HPC strategies}.

Regarding mixed-precision, the clean-up sweeps may use a mixed-precision diagonalization method, which gives rise to a two-level mixed-precision strategy~\cite{Tian2022}. However, we do not make this distinction in \Cref{tab: HPC strategies}.

\subsection{Symmetries}

 Next, we elaborate on the symmetry aspects introduced in \Cref{ssec: High-Level Overview}. As before, we classify packages based on whether they support each symmetry in any form, without distinguishing between those whose support is limited to specific built-in models and those that allow them to be arbitrarily combined. \Cref{tab: symmetry} indicates support for the following symmetries:
\begin{itemize}
    \item \emph{The unitary group $\mathrm{U}(1)$:} Indicates support for the unitary group of degree one, which is the most commonly used symmetry. It typically enforces the conservation of particle numbers (total electrons or separate spin counts) and total spin magnetization.
    
    \item \emph{Cyclic groups {\rm ($\mathbb{Z}$)}:} Indicates support for either $\mathbb{Z}_2$ or more general $\mathbb{Z}_n$ symmetries. A common usage for $\mathbb{Z}_2$ is to enforce the spin-flip (parity) symmetry in the transverse-field Ising model, where flipping all spins leaves the Hamiltonian unchanged. The more general $\mathbb{Z}_n$ group is common in models with an $n$-fold rotational or cyclic invariance.

    \item \emph{Special unitary groups {\rm ($\mathrm{SU}$)}:} Indicates support for either $\mathrm{SU}(2)$ or more generally $\mathrm{SU}(n)$. The $\mathrm{SU}(2)$ group is most common and is often applied to preserve the total spin, often referred to as spin $\mathrm{SU}(2)$. An example of $\mathrm{SU}(n)$ is the channel $\mathrm{SU}(3)$ symmetry considered by Weichselbaum~\cite{Weichselbaum2012}. 
    
    \item \emph{Point groups {\rm ($\mathrm{P}$)}:} Indicates support for abelian point group symmetries $\mathrm{P} \in \{C_1, \allowbreak C_i,\allowbreak C_2,\allowbreak C_s,\allowbreak C_{2h},\allowbreak D_2,\allowbreak C_{2v},\allowbreak D_{2h}\}$ with real-valued character tables. They are often used for molecular systems to enforce spatial symmetries~\cite{Wouters2014Density}.

    \item \emph{Fermion parity {\rm ($\mathbb{Z}_2^\mathrm{f}$)}:} Indicates support for the fermion parity, which captures the anti-commuting nature of fermionic degrees of freedom. The parity is a $\mathbb{Z}_2$ quantum number that yields an additional sign when exchanging two fermions (each carrying odd parity).

    \item \emph{Other:} Indicates support for less common symmetries, such as the special orthogonal group $\mathrm{SO}(n)$, the symplectic group $\mathrm{Sp}(2n)$, and Anyonic symmetries \cite{Weichselbaum2012}. Anyons are particles with non-trivial exchange statistics that are neither fermions nor bosons.
\end{itemize}

\begin{table}[htbp]
    \centering
    \caption{Support for various types of symmetries. Note that support for multiple features does not imply that they can be utilized simultaneously.}
    \begin{tabular}{llcccccl}
\toprule
{\bf ID} & {\bf Name} & {\bf $\bm{\mathrm{U}(1)}$} & {\bf $\bm{\mathbb{Z}}$} & {\bf $\bm{\mathrm{SU}}$} & {\bf $\bm{\mathrm{P}}$} & {\bf $\bm{\mathbb{Z}_2^\mathrm{f}}$} & {\bf Other} \\
\midrule
1 & ALPS DMRG & \checkmark & - & - & - & \checkmark & - \\ \hline
2 & ALPS MPS & \checkmark & $2$ & - & - & \checkmark & - \\ \hline
3 & BAGEL & \checkmark & - & - & - & \checkmark & - \\ \hline
4 & Block2 & \checkmark & $n$ & $2$ & \checkmark & \checkmark & - \\ \hline
5 & CheMPS2 & \checkmark & - & $2$ & \checkmark & \checkmark & - \\ \hline
6 & ChemTensor & \checkmark & - & $2$ & - & - & - \\ \hline
7 & Chen et al. & \checkmark & - & - & - & \checkmark & - \\ \hline
8 & Cytnx & \checkmark & $n$ & - & - & \checkmark & - \\ \hline
9 & DMRG-Budapest & \checkmark & $n$ & $n$ & \checkmark & \checkmark & - \\ \hline
10 & DMRG++ & \checkmark & - & - & - & \checkmark & - \\ \hline
11 & DMRGPy & - & - & - & - & \checkmark & - \\ \hline
12 & FOCUS & \checkmark & - & $2$ & - & \checkmark & - \\ \hline
13 & Hong et al. & - & - & - & - & - & - \\ \hline
14 & ITensor & \checkmark & $n$ & - & - & \checkmark & - \\ \hline
15 & ITensorMPS.jl & \checkmark & $n$ & - & - & \checkmark & - \\ \hline
16 & Kylin & \checkmark & - & $2$ & \checkmark & \checkmark & - \\ \hline
17 & MOLMPS & \checkmark & - & - & \checkmark & \checkmark & - \\ \hline
18 & MPSKit.jl & \checkmark & $n$ & $n$ & - & \checkmark & Anyonic/Categoric \\ \hline
19 & MPToolkit & \checkmark & $n$ & $2$ & - & \checkmark & - \\ \hline
20 & OSMPS & \checkmark & $2$ & - & - & \checkmark & - \\ \hline
21 & PyTeNet & \checkmark & - & - & - & - & - \\ \hline
22 & pyTTN & - & - & - & - & \checkmark & - \\ \hline
23 & QCMaquis & \checkmark & $2$ & $2$ & \checkmark & \checkmark & - \\ \hline
24 & QSpace & \checkmark & $n$ & $n$ & - & \checkmark & SO(n), Sp(2n) \\ \hline
25 & Quantum TEA & \checkmark & $n$ & - & - & \checkmark & - \\ \hline
26 & quimb & - & - & - & - & \checkmark & - \\ \hline
27 & Renormalizer & \checkmark & - & - & - & \checkmark & - \\ \hline
28 & SeeMPS2 & - & - & - & - & - & - \\ \hline
29 & SUNDMRG.jl & - & - & $n$ & - & - & - \\ \hline
30 & SymMPS & \checkmark & - & - & - & \checkmark & - \\ \hline
31 & SyTen & \checkmark & $n$ & $2$ & - & \checkmark & - \\ \hline
32 & TeNPy & \checkmark & $n$ & - & - & \checkmark & - \\ \hline
33 & tensor-tools & \checkmark & - & - & - & \checkmark & - \\ \hline
34 & TensorTrack & \checkmark & $n$ & $n$ & - & \checkmark & - \\ \hline
35 & UltraDMRG & \checkmark & $n$ & - & - & \checkmark & - \\ \hline
36 & xDMRG++ & - & - & - & - & \checkmark & - \\ \hline
37 & YASTN & \checkmark & $n$ & - & - & \checkmark & - \\ 
\bottomrule
\end{tabular}

    \label{tab: symmetry}
\end{table}

The fermion parity is often implemented using the Jordan-Wigner transformation, which maps fermionic creation and annihilation operators to (bosonic) spin operators. However, newer approaches incorporate it directly in tensor networks on arbitrary graphs using graded Hilbert spaces~\cite{Bultinck2017, Mortier2025, Gao2024}.

Finally, some software packages support user-defined symmetry groups, where the user specifies the properties/rules of the symmetry by providing appropriate Clebsch-Gordan coefficients or encoding them via an abstract tensor interface available in the package, for example. However, this type often requires complex workflows and a deep understanding of the software, as the interfaces tend to be complex. Additionally, it can be challenging to categorize and determine the threshold for what qualifies as supporting user-defined symmetries. Hence, we excluded them from \Cref{tab: symmetry} to maintain clarity and consistency.

We conclude that $\mathrm{U}(1)$ and $\mathbb{Z}_2^\mathrm{f}$ are the most frequently supported forms of symmetries.
Support for non-abelian symmetries is mainly restricted to $\mathrm{SU}(2)$. 

\subsection{Hamiltionians}

We investigate the types of Hamiltonians available in the selected packages. We try to provide broad insight into each package's flexibility in defining system interactions. However, the classifications in this section are somewhat up for interpretation due to variations in perspective. \Cref{tab: hamiltonian} presents the following information:
\begin{itemize}
    \item \emph{Construction:} Indicates if and how a user can define custom Hamiltonians. 
    \begin{itemize}
        \item \emph{Custom:} Indicates support for some generic custom Hamiltonian construction interface. 

        \item \emph{Operator:} Indicates support for constructing custom Hamiltonians using local $n$-body operators defined on the different site-indices upon which the operators act. 
    \end{itemize}
    
    \item \emph{Built-in:} Indicates whether the package contains built-in Hamiltonians that target a specific or broad range of fields.
\end{itemize}

\begin{table}[hbtp]
    \centering
    \caption{Support for custom Hamiltonian constructions and the variety of fields for built-in models.}
    \begin{tabular}{llccl}
\toprule
\multirow{2}{*}{{\bf ID}} & \multirow{2}{*}{{\bf Name}} & \multicolumn{2}{c}{{\bf Construction}} & \multirow{2}{*}{{\bf Built-in}} \\
\cmidrule(lr){3-4}
 &  & {\bf Custom} & {\bf Operator} &  \\
\midrule
1 & ALPS DMRG & \checkmark & \checkmark & Specific \\ \hline
2 & ALPS MPS & - & - & Specific \\ \hline
3 & BAGEL & - & - & Specific \\ \hline
4 & Block2 & \checkmark & \checkmark & Broad \\ \hline
5 & CheMPS2 & - & - & Specific \\ \hline
6 & ChemTensor & \checkmark & \checkmark & Broad \\ \hline
7 & Chen et al. & - & - & Broad \\ \hline
8 & Cytnx & \checkmark & \checkmark & - \\ \hline
9 & DMRG-Budapest & \checkmark & \checkmark & Broad \\ \hline
10 & DMRG++ & \checkmark & - & Specific \\ \hline
11 & DMRGPy & \checkmark & \checkmark & - \\ \hline
12 & FOCUS & - & - & Specific \\ \hline
13 & Hong et al. & - & - & Specific \\ \hline
14 & ITensor & \checkmark & \checkmark & - \\ \hline
15 & ITensorMPS.jl & \checkmark & \checkmark & - \\ \hline
16 & Kylin & - & - & Specific \\ \hline
17 & MOLMPS & - & - & Broad \\ \hline
18 & MPSKit.jl & \checkmark & \checkmark & Specific \\ \hline
19 & MPToolkit & \checkmark & - & Broad \\ \hline
20 & OSMPS & \checkmark & \checkmark & - \\ \hline
21 & PyTeNet & \checkmark & \checkmark & Broad \\ \hline
22 & pyTTN & \checkmark & \checkmark & Broad \\ \hline
23 & QCMaquis & - & - & Broad \\ \hline
24 & QSpace & \checkmark & \checkmark & Specific \\ \hline
25 & Quantum TEA & \checkmark & \checkmark & Specific \\ \hline
26 & quimb & \checkmark & \checkmark & Specific \\ \hline
27 & Renormalizer & \checkmark & \checkmark & Broad \\ \hline
28 & SeeMPS2 & \checkmark & \checkmark & - \\ \hline
29 & SUNDMRG.jl & - & - & Specific \\ \hline
30 & SymMPS & \checkmark & - & - \\ \hline
31 & SyTen & \checkmark & - & - \\ \hline
32 & TeNPy & \checkmark & \checkmark & Broad \\ \hline
33 & tensor-tools & \checkmark & \checkmark & - \\ \hline
34 & TensorTrack & \checkmark & \checkmark & Specific \\ \hline
35 & UltraDMRG & \checkmark & \checkmark & - \\ \hline
36 & xDMRG++ & - & - & Specific \\ \hline
37 & YASTN & \checkmark & \checkmark & - \\ 
\bottomrule
\end{tabular}

    \label{tab: hamiltonian}
\end{table}

To provide further clarity regarding \Cref{tab: hamiltonian}, firstly, we only indicate the presence of some interface for custom Hamiltonian construction and do not distinguish based on their implementation. Moreover, standard operators include spin, spin exchange, bosonic, and fermionic operators for the operator-type construction interfaces. But again, we do not make distinctions based on how many or which operators each package offers.

Secondly, we acknowledge that classifying the variety of built-in Hamiltonians is somewhat subjective. Currently, the column reflects the developers' views on their package. We provided them with the following examples of what we considered fields: quantum chemistry, condensed matter physics, and nuclear structure.
\subsection{Eigensolvers}

We investigate whether the packages utilize an external dependency for their eigensolver or if they implement their own. \Cref{tab: eigensolvers} shows:
\begin{itemize}
    \item \emph{Eigensolver:} Indicates if the eigensolver is a bespoke implementation or provided by some external dependency. Some packages include both, often implementing their own sparse solver complemented by an external dense solver.
    
    \item \emph{Comment:} Short comment regarding the method used, mainly for packages that implement their own eigensolver, as they can be less flexible regarding method changes. 
\end{itemize}

\begin{table}[htbp]
    \centering
    \caption{The eigensolvers used by each package and a comment for bespoke implementations.}
    \begin{tabular}{llll}
\toprule
{\bf ID} & {\bf Name} & {\bf Eigensolver} & {\bf Comment} \\
\midrule
1 & ALPS DMRG & Own & Lanczos \\ \hline
2 & ALPS MPS &  IETL in ALPS & Jacobi–Davidson \\ \hline
3 & BAGEL & Own & Davidson \\ \hline
4 & Block2 & Own & Davidson \\ \hline
5 & CheMPS2 & Own & Davidson \\ \hline
6 & ChemTensor & Own & Lanczos \\ \hline
7 & Chen et al. & Own & Davidson \\ \hline
8 & Cytnx & Own & Lanczos \\ \hline
9 & DMRG-Budapest & Own, ARPACK & Lanczos, Davidson, etc. \\ \hline
10 & DMRG++ & Own & Lanczos \\ \hline
11 & DMRGPy & ITensor (\CC/Julia) & - \\ \hline
12 & FOCUS & Own & Davidson \\ \hline
13 & Hong et al. & Own & Lanczos \\ \hline
14 & ITensor & Own & Davidson \\ \hline
15 & ITensorMPS.jl & KrylovKit.jl & Lanczos, Arnoldi \\ \hline
16 & Kylin & Own & Lanczos, Davidson \\ \hline
17 & MOLMPS & Own & Davidson \\ \hline
18 & MPSKit.jl & KrylovKit.jl & Lanczos, Arnoldi \\ \hline
19 & MPToolkit & ARPACK & Lanczos, Davidson, etc. \\ \hline
20 & OSMPS & Own, ARPACK & Lanczos \\ \hline
21 & PyTeNet & Own & Lanczos \\ \hline
22 & pyTTN & Own & Arnoldi \\ \hline
23 & QCMaquis & ALPS & Jacobi–Davidson \\ \hline
24 & QSpace & Own & Davidson \\ \hline
25 & Quantum TEA & Own, SciPy & Lanczos \\ \hline
26 & quimb & SciPy, slepc4py & Krylov-Schur, Davidson, etc. \\ \hline
27 & Renormalizer & Own, SciPy & Davidson \\ \hline
28 & SeeMPS2 & SciPy & Lanczos \\ \hline
29 & SUNDMRG.jl & Own & Lanczos \\ \hline
30 & SymMPS & Own & Lanczos \\ \hline
31 & SyTen & Own & Lanczos, Davidson \\ \hline
32 & TeNPy & Own, SciPy & Lanczos \\ \hline
33 & tensor-tools & Own & Davidson \\ \hline
34 & TensorTrack & Own & Krylov-Schur \\ \hline
35 & UltraDMRG & Own & Lanczos \\ \hline
36 & xDMRG++ & PRIMME & Arnoldi, Davidson, etc. \\ \hline
37 & YASTN & Own & Lanczos \\ 
\bottomrule
\end{tabular}

    \label{tab: eigensolvers}
\end{table}

We conclude that most packages include a custom implementation of the eigensolver component. 
Most implementations are based on either the Lanczos or the Davidson algorithms.

\section{Conclusion} 
\label{sec: Conclusion}

We systematically surveyed the software landscape for the DMRG algorithm and identified more than 50 packages, 37 of which we compared in multiple aspects of their features. 

The abundance of packages raises questions about why so many were created and why so many are still actively maintained.
We found that one explanation is that most packages were developed by a small team around a research group to address specific needs related to their applications.  
What may have started as a small proof-of-concept gradually grows into a sophisticated package over many years. 
The flexibility granted by being in complete control over a package has clear advantages. 
While this could explain an initial surge of packages to some extent, it does not adequately explain why we have not yet seen a consolidation around a few projects supported by larger communities. 
The survey suggests that one reason is the inherently complicated nature of the input.
There are many alternative ways to specify a Hamiltonian and symmetries, none of which seems to be superior. 
Consequently, DMRG implementations are rarely interchangeable, and migration requires a considerable up-front development cost. 
The survey also revealed a great variety of supported feature combinations. 

An immediate drawback of redundant implementations of core components is the need to reimplement new techniques in each package.
Another disadvantage is that specialized packages may have a design tailored to their intended use cases. 
They may therefore be difficult to adapt when new techniques are discovered. 
This problem becomes particularly acute when dealing with complex challenges such as adding GPU support, non-abelian symmetries, adaptive load balancing, or auto-tuning. 

Reducing the number of independent implementations translates into lower overall maintenance costs and a more immediate spread of the benefits of new advances.
All packages rely on BLAS instead of re-implementing matrix multiplication.
A similar modular approach could potentially be applied successfully to other operations common to most implementations.
Possible targets for modularization include tensor operations, symmetry representations, and eigensolvers.
Another strategy to reduce the proliferation of packages is for similar projects to merge into larger community projects.

Some recent efforts, such as the modular designs of \textsc{ITensors.jl}~\cite{ITensors.jlSourceCode}, \textsc{TensorKit.jl} \cite{TensorKit.jlSourceCode}, and \textsc{quimb}~\cite{quimbSourceCode}, and the collaborative effort to combine \textsc{Cytnx}~\cite{CytnxSourceCode} and \textsc{TeNPy} \cite{TeNPySourceCode} into \textsc{Cyten}~\cite{CytenSourceCode} point to a growing recognition of the benefits of modularization and collaborative development. 
In fact, several current efforts specifically aim at bringing developers of DMRG and related software together. 
Examples include the Reusable Libraries in Quantum Chemistry meeting~\cite{HelsinkiConference_web} and the Toulouse Tensor Workshop~\cite{ToulouseTensorWorkshop_web}

We believe that the reason we see so many different implementations is more social than technical.
Our ambition is for our review to raise awareness of the DMRG software landscape and inspire more inter-project and inter-field collaboration around the bottom-up modularization of common functionality and the top-down consolidation of similar projects.

\section*{Acknowledgements} 

We thank the developers of the surveyed DMRG packages who took the time to answer our questions about their packages, as an essential part of this work has been based on the information they provided.

Supported by the eSSENCE Programme under the Swedish Government’s Strategic Research Initiative.

J. Brandejs has received funding from the European Research Council (ERC) 
under the European Union’s Horizon 2020 research and innovation 
programme (grant agreement No 101019907).

\bibliographystyle{elsarticle-num} 
\bibliography{references, references-packages}

\end{document}